\documentstyle[aps,prl,epsf]{revtex}
\begin{document}

\title{Magnetization reversal by uniform rotation \\
(Stoner--Wohlfarth model) in f.c.c. cobalt nanoparticles}

\author{W. Wernsdorfer$^{1}$, C. Thirion$^{1}$, N. Demoncy$^{2}$, 
         H. Pascard$^{2}$, D. Mailly$^{3}$}

\address{
$^{1}$ Laboratoire Louis N\'eel-CNRS, BP 166,38042 Grenoble, France.\\
$^{2}$ SESI, CEA/CNRS, Ecole Polytechnique, 91128 Palaiseau, France\\ 
$^{3}$ LPN, CNRS, 196 avenue H. Ravera, 92220 Bagneux, France.
}
\date{Version: 3 june 2001}
\maketitle

\begin{abstract}
The combination of high sensitive superconducting quantum interference 
device (SQUID) with high quality nanoparticles allowed to check 
the simplest classical model describing the magnetisation reversal
by uniform rotation which were proposed more than 50 years 
ago by N\'eel, Stoner and Wohlfarth. The micrometer sized SQUIDs 
were elaborated by electron beam lithography and the nanoparticles 
were synthesised by arc-discharge. 
The measured angular dependence of switching
fields of nearly all f.c.c. Co nanoparticles revealed a
dominating uniaxial magnetic anisotropy.
This result suggests that twin boundaries and 
stacking faults strongly alter the
cubic magnetocrystalline anisotropy leading to dominating uniaxial anisotropy.
However, few particles were sufficiently "perfect"
in order to show a more complex switching
field surface and a field path dependence of the switching
field which is the important 
signature of the cubic magnetocrystalline anisotropy. 
\end{abstract}

\bigskip

For a sufficiently small magnetic sample it 
is energetically unfavourable to form a stable magnetic domain wall. 
The specimen then behaves as a single magnetic domain~\cite{Aharoni96}. 
For the smallest single-domain particles, the 
magnetisation is expected to 
reverse by uniform rotation of magnetisation. 
For somewhat larger ones, nonuniform reversal modes 
are more likely--for example, the curling 
reversal mode. 
In this paper, we discuss in detail the uniform rotation 
mode that is used in many theories, 
in particular in N\'eel, Brown, and 
Coffey's theory of magnetisation reversal by 
thermal activation~\cite{Coffey98a} 
and in the theory of 
macroscopic quantum tunnelling of magnetisation~\cite{QTM94}.
For a review, see Ref. \cite{WW_ACP_01}.

\section{Generalisation of the Stoner--Wohlfarth model}

The model of uniform rotation of magnetisation, developed by 
Stoner and Wohlfarth~\cite{St_W48} and N\'eel~\cite{Neel47}, 
is the simplest classical model describing 
magnetisation reversal. One considers a particle of an 
ideal magnetic material where exchange 
energy holds all spins tightly parallel to each other, and the 
magnetisation magnitude does not depend 
on space. In this case the exchange energy 
is constant, and it plays no role in the energy 
minimisation. Consequently, there is competition only 
between the anisotropy energy of the 
particle and the effect of the applied field. 

The original model of Stoner and Wohlfarth assumed only uniaxial 
shape anisotropy with one anisotropy constant---that is, 
one second-order term. 
This is sufficient to describe highly symmetric cases like a prolate 
spheroid of revolution or an infinite cylinder. However, real systems 
are often quite complex, and the anisotropy is a sum of mainly shape 
(magnetostatic), magnetocrystalline, magnetoelastic, and surface anisotropy. 
One additional 
complication arises because the different contributions of 
anisotropies are often aligned in an arbitrary way one
with respect to each other. 
All these facts motivated a generalisation of the Stoner--Wohlfarth 
model for an arbitrary effective anisotropy which was done by 
Thiaville~\cite{Thiaville98,Thiaville00}. 

Similar to the Stoner--Wohlfarth model, one
supposes that the exchange interaction in the particle
couples all the spins strongly together to form a
giant spin whose direction is described by the unit vector $\vec{m}$.
The only degrees of freedom of the particle's magnetisation are the
two angles of orientation of $\vec{m}$.
The reversal of the magnetisation is described by the
potential energy
\begin{equation}
E(\vec{m},\vec{H}) = E_0(\vec{m}) - \mu_0 V M_{\rm s} \vec{m}.\vec{H}
\label{eq_E}
\end{equation}
where $V$ and $M_{s}$ are the magnetic volume
and the saturation magnetisation of the particle respectively, 
$\vec{H}$ is
 the external magnetic field,
and $E_0(\vec{m})$ the magnetic anisotropy energy which is given by
\begin{equation}
E_0(\vec{m}) = E_{\rm shape}(\vec{m})
	+ E_{\rm MC}(\vec{m})
	+ E_{\rm surface}(\vec{m})
	+ E_{\rm ME}(\vec{m})
\label{eq_E_0}
\end{equation}

$E_{\rm shape}$ is the magnetostatic energy related to
the particle shape. 
$E_{\rm MC}$ is the magnetocrystalline anisotropy (MC) 
arising from the coupling
of the magnetisation with the crystalline 
lattice, similar as in bulk.
$E_{\rm surface}$ is due to the symmetry breaking and
surface strains.
In addition, if the particle experiences an external stress, the 
volumic
relaxation inside the particle induces a magnetoelastic (ME) 
anisotropy energy $E_{\rm ME}$.
All these anisotropy energies can be developed in a power
series of $m_{x}^{a}m_{y}^{b}m_{z}^{c}$ 
with $p = a + b + c = 2, 4, 6,\ldots$ 
giving the order of the anisotropy term. 
Shape anisotropy can be written as a biaxial anisotropy with two 
second-order terms. 
Magnetocrystalline anisotropy is in most 
cases either uniaxial (hexagonal systems) or cubic, yielding 
mainly second- and fourth-order terms. Finally, in the simplest 
case, surface and magnetoelastic anisotropies are of second order. 

Thiaville proposed a geometrical method 
to calculate the particle's energy and
to determine the switching field for all angles of the applied 
magnetic field yielding the critical surface of switching fields
which is analogous to the Stoner--Wohlfarth astroid.

The main interest of Thiaville's calculation is that measuring 
the critical surface of the switching field allows one to find 
the effective anisotropy of the nanoparticle. The knowledge of 
the latter is important for temperature-dependent studies 
and quantum tunnelling investigations. 
Knowing precisely the particle's shape and the crystallographic axis 
allows one to determine the different contributions to the 
effective anisotropy.

\section{Uniform Rotation with cubic anisotropy}

We have seen in the previous 
studies~\cite{WW_PRL97_Co,WW_PRL97_BaFeO,Bonet99,Jamet01a} 
that the magnetic anisotropy is 
often dominated by second-order anisotropy terms. However, for 
nearly symmetric shapes, fourth-order terms 
\footnote{For example, the fourth-order terms of {\it f.c.c.} 
magnetocrystalline anisotropy.} can be 
comparable or even dominant with respect to the second-order terms. 
Therefore, it is interesting to discuss further the features of fourth 
order terms. We restrict the discussion to the 2D 
problem~\cite{Thiaville98,Ching91} (see Ref.~\cite{Thiaville00} for 3D).

The reversal of the magnetisation is described by Eq.~(\ref{eq_E}) 
that can be rewritten in 2D
\begin{equation}
E(\theta) = E_0(\theta) - 
               \mu_0 v M_{\rm s} (H_x \cos(\theta) + H_y \sin(\theta))
\label{eq_E_2D}
\end{equation}
where $v$ and $M_{s}$ are the magnetic volume
and the saturation magnetisation of the particle, respectively, $\theta$ is
the angle between the magnetisation direction and $x$, $H_x$ and $H_y$ 
are the components of the external magnetic field along $x$ and $y$,
and $E_0(\theta)$ is the magnetic anisotropy energy.
The conditions of critical fields ($\partial E/\partial \theta = 0$ and 
$\partial E^2/\partial \theta^2 = 0$) yield a parametric form of the locus of 
switching fields
\begin{eqnarray}
	H_x = - \frac{1}{2 \mu_0 v M_{\rm s}} 
	        \left( \sin(\theta) \frac{dE}{d\theta} +
			       \cos(\theta) \frac{d^2E}{d\theta^2} \right)
	\label{eq_hx}
\\
	H_y = + \frac{1}{2 \mu_0 v M_{\rm s}} 
	        \left( \cos(\theta) \frac{dE}{d\theta} -
			       \sin(\theta) \frac{d^2E}{d\theta^2} \right) 
	\label{eq_hy}
\end{eqnarray}

\begin{figure}
\begin{center}\leavevmode
\centerline{\epsfxsize=8 cm \epsfbox{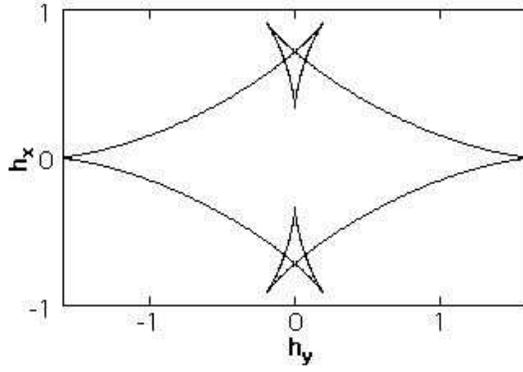}}
\caption{Angular dependence of the 
switching field obtained from 
Eqs.~(\ref{eq_hx})--(\ref{eq_E0_fcc}) with $K_1 > 0$ 
and $K_2 = - 2/3 K_1$. The field is 
normalised by the factor $2 K_1 / (\mu_0 M_{\rm s})$.}
\label{ast_fcc}
\end{center}
\end{figure}

\begin{figure}
\begin{center}\leavevmode
\centerline{\epsfxsize=8 cm \epsfbox{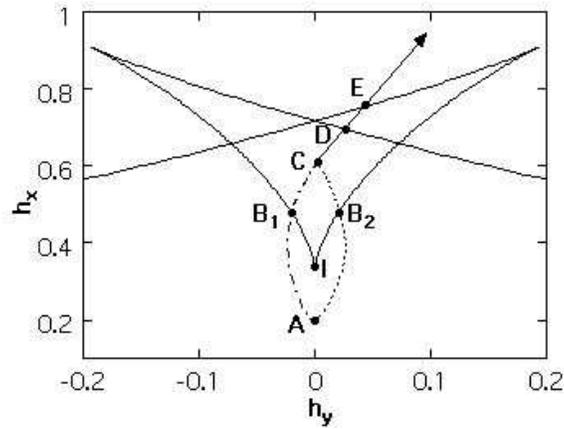}}
\caption{Enlargement of angular dependence 
of the switching field of 
Fig.~\ref{ast_fcc}. Two possible paths 
of the applied field are 
indicated: Starting from point {\bf A} and going 
over the point {\bf B$_1$} leads to magnetisation reversal 
in {\bf E}, whereas going over the point {\bf B$_2$} leads to reversal 
in {\bf D}.}
\label{ast_fcc_zoom}
\end{center}
\end{figure}

\begin{figure}
\begin{center}\leavevmode
\centerline{\epsfxsize=7 cm \epsfbox{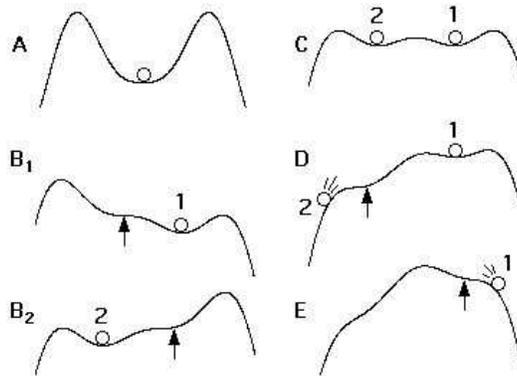}}
\caption{Scheme of the potential energy near the metastable state 
for different applied fields as indicated in Fig.~\ref{ast_fcc_zoom}.
The balls represent the state of the magnetisation, and the arrows 
locate the appearing or disappearing well.}
\label{potential_fcc}
\end{center}
\end{figure}

As an example we study a system with uniaxial shape 
anisotropy and cubic anisotropy.
The total magnetic anisotropy energy can be described by
\begin{equation}
E_0(\theta) = v K_1 \sin^2(\theta + \theta_0) + 
              v K_2 \sin^2(\theta)\cos^2(\theta) 
\label{eq_E0_fcc}
\end{equation}
where $K_1$ and $K_2$ are anisotropy constants ($K_1$ could be a 
shape anisotropy and $K_2$ the cubic crystalline anisotropy of a {\it 
f.c.c.} crystal.) $\theta_0$ is a constant which allows to turn one 
anisotropy contribution with respect to the other one.
Fig.~\ref{ast_fcc} displays an example of a critical curve 
which can easily be calculated from 
Eqs.~(\ref{eq_hx})--(\ref{eq_E0_fcc}).
When comparing the standard Stoner--Wohlfarth astroid 
with Fig.~\ref{ast_fcc}, we can realise that the 
critical curve can cross itself several times. In this case, the 
switching field of magnetisation depends on 
the path followed by the applied 
field. In order to understand this point, let us follow the energy 
potential [Eqs.~(\ref{eq_E_2D}) and (\ref{eq_E0_fcc})] when sweeping the 
applied field as indicated in Fig.~\ref{ast_fcc_zoom}. When the field 
is in {\bf A}, the energy $E$ has two minima and the magnetisation is in 
the metastable potential well. As the field increases, the metastable 
well becomes less and less stable. Let us compare two paths, one 
going along {\bf A} $\rightarrow$ {\bf B$_1$} $\rightarrow$ 
{\bf C} $\rightarrow$ {\bf D} $\rightarrow$ {\bf E}, 
the other over {\bf B$_2$} instead of {\bf B$_1$}.
Fig.~\ref{potential_fcc} shows $E$ in the vicinity of the 
metastable well for different field values along 
the considered paths (the stable potential well
is not presented).
One can realise that the state of the 
magnetisation in {\bf C} depends on 
the path followed by the field: Going over 
{\bf B$_1$} leads to the magnetisation state in the 
left metastable well (1), whereas going 
over {\bf B$_2$} leads to the right 
metastable well (2). The latter path leads to magnetisation switching 
in {\bf D}, and the former one leads to a switching in {\bf E}. Note that a small
magnetisation switch happens when reaching {\bf B$_1$} or {\bf B$_2$}.
Point {\bf I} is a supercritical bifurcation.

\newpage

\section{Experimental evidence for magnetisation 
reversal by uniform rotation}

In order to demonstrate experimentally the 
uniform rotation mode, the angular 
dependence of the magnetisation reversal has often been studied 
(see references in Ref.~\cite{Aharoni96}). However, a comparison of 
theory with experiment is difficult because magnetic 
particles often have a nonuniform magnetisation 
state that is due to rather complicated 
shapes and surfaces, crystalline defects, and surface anisotropy. In 
general, for many 
particle shapes the demagnetisation fields inside the particles are 
nonuniform leading to 
nonuniform magnetisation states~\cite{Aharoni96}. 

An example are ultrathin magnetic dots 
with in-plane uniaxial anisotropy 
showing switching fields that are very 
close to the Stoner--Wohlfarth model, although magnetic 
relaxation experiments clearly showed
that nucleation volumes are by far smaller than an 
individual dot volume~\cite{Fruchart99_Fe} being
in agreement with calculations~\cite{Fruchart01}. 
These studies show clearly that switching field measurements 
as a function of the angles of the applied field cannot be
taken unambiguously as a proof of a Stoner--Wohlfarth reversal.

The first clear demonstration of the uniform reversal mode has been
found with Co nanoparticles~\cite{WW_PRL97_Co}, and 
BaFeO nanoparticles~\cite{WW_PRL97_BaFeO}, 
the latter having a dominant uniaxial 
magnetocrystalline anisotropy. 
The three-dimensional angular dependence
of the switching field measured on BaFeO 
particles of about 20~nm could be explained with the
Stoner--Wohlfarth model taking into account the shape 
anisotropy and hexagonal crystalline anisotropy of 
BaFeO~\cite{Bonet99}. 
This explication were supported by temperature- and 
time-dependent measurements yielding activation volumes
which are very close to the particle volume.
More recently, the three-dimensional switching
field measurements of individual 3 nm Co nanoparticles
showed also the uniform reversal mode with dominant 
uniaxial magnetocrystalline anisotropy which was dominated
by surface anisotropy (particle--matrix interface)~\cite{Jamet01a}. 

The study of uniform rotation for particles with dominating
cubic anisotropy revealed to be very difficult because
nearly all structural defects in particles lead to 
dominating uniaxial anisotropy. Nevertheless, we could
find few particles which were sufficiently "perfect"
in order to show clearly a field path dependence of the switching
field (Fig.~\ref{ast_fcc}) which is the important 
signature of strong higher order terms in the 
potential energy (Eq.~(\ref{eq_E}).
The first measurement of such a field path dependence of switching 
fields were performed on single-domain FeCu nanoparticles of about 
15 nm with a cubic crystalline anisotropy and a small 
arbitrarily oriented shape anisotropy~\cite{Bonet98}.
Here we report on the magnetic study of individual 
single-domain cobalt nanoparticles encapsulated 
in a carbon envelope which provides a very efficient protection 
against oxidation. Our results show an experimental agreement 
with the simple classical model mentioned above.

\begin{figure}
\begin{center}\leavevmode
\centerline{\epsfxsize=9 cm \epsfbox{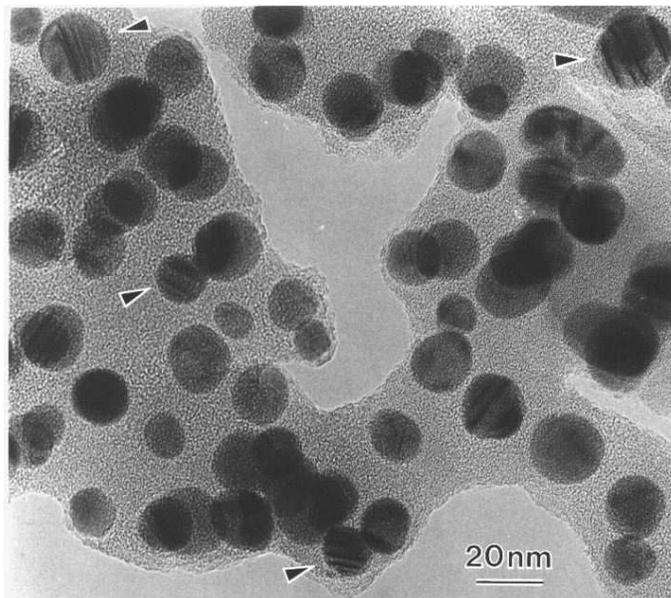}}
\caption{TEM image of an assembly of typical cobalt 
nanoparticles encapsulated in a few graphitic sheets 
and a large amount of amorphous carbon. Most of the particles 
have a spherical shape checked through tilting experiments 
and frequently contain planar defects (see arrows).}
\label{SEM_Co_1}
\end{center}
\end{figure}

\begin{figure}
\begin{center}\leavevmode
\centerline{\epsfxsize=9 cm \epsfbox{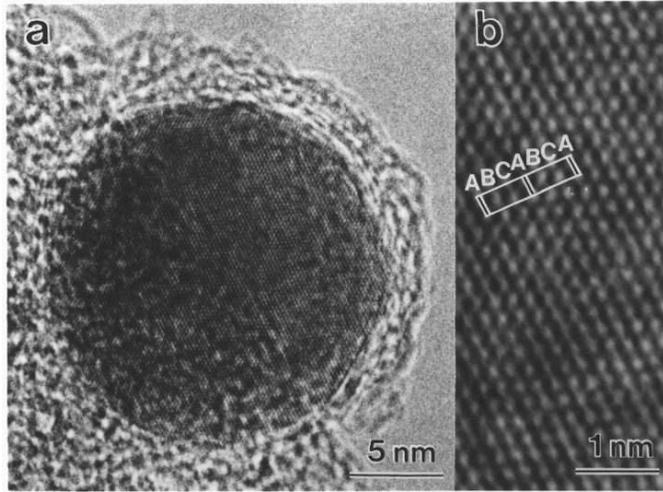}}
\caption{(a) HRTEM image of a typical single-crystalline 
f.c.c.--Co nanoparticle in $<$110$>$ projection. 
Only 3 or 4 graphite sheets imaged by $<$002$>$ lattice fringes 
with a separation of 0.34 nm are stacked parallel to its surface. 
Amorphous carbon surrounds the graphite envelope. 
(b) enlargement of (a) (digitised and Fourier filtered image) 
showing the 'ABC' type stacking of $<$111$>$ dense atomic 
layers which is characteristic of the f.c.c. structure. 
The slight deviation we noted in angles measurements is 
attributed to distortions of the structure induced by the interface 
between the nanocrystal and its graphitic envelope.}
\label{SEM_Co_2}
\end{center}
\end{figure}

\subsection{Synthesis of nanoparticles}

The arc-discharge method using two graphite electrodes has 
initially been developed for the synthesis of 
fullerenes~\cite{Kroto85,Kratschmer90} and carbon 
nanotubes~\cite{Iijima91} and has been then adapted 
to the production of endohedral fullerenes and carbon 
coated nanocrystallites. This technique was used to synthesise 
the cobalt nanoparticles encapsulated in carbon: the graphite 
anode was drilled and packed with a mixture of 
graphite and pure cobalt powders~\cite{Guerret94}. 
A plasma was established in the helium atmosphere under 
the following conditions : 0.6 bar, 25-30 V, 100-105 A dc, 
during 45 minutes. The deposit formed on the graphite 
cathode was ground, ultrasonically dispersed in ethanol 
and put on a holey carbon film for Transmission Electron 
Microscopy (TEM).

The cobalt nanoparticles are very abundant whereas 
only a few partially filled carbon nanotubes can be found. 
Most of the nanocrystals have a spherical shape with a 
diameter ranging from 5 to 30 nm (Fig.~\ref{SEM_Co_1}). 
All the particles are coated by carbon. 
For particles smaller than 10 nm in diameter, 
the carbon coating is amorphous whereas 
large particles are encapsulated in graphitic 
cages and their shape is more polyhedral. 
Most of the particles have only 3 or 4 graphitic sheets 
stacked parallel to their surface and are surrounded 
with amorphous carbon. No void has been observed between 
the crystallite and the carbon envelope which avoids 
the oxidation on the surface of the particles 
that could dramatically affect the magnetic properties.

The crystallites are pure cobalt and have a face-centered 
cubic (f.c.c.) structure according to the electron 
diffraction patterns obtained from assemblies of particles. 
However, the presence of a small amount of hexagonal 
close-packed (h.c.p.) particles cannot be excluded. 
Fig.~\ref{SEM_Co_2} shows an High Resolution TEM image of a 
typical f.c.c.-Co nanoparticle in a $<$110$>$ projection 
and encapsulated in 3 or 4 graphitic sheets. 
No evidence of cobalt carbides has been found. 
This indicates that the particles were formed at high 
temperature and rapidly quenched since f.c.c.-Co is 
the high temperature phase and h.c.p.-Co and cobalt 
carbides are not stable above 500$^{\circ}$C. The nanoparticles 
are mostly single-crystals and often contain 
twin boundaries and stacking faults (Fig.~\ref{SEM_Co_1}) since 
these planar defects are known to occur frequently 
in f.c.c.-Co. As revealed by the HRTEM images, 
the surface roughness generally does not exceed 
2 or 3 atomic layers : this is another important point 
for the quality of particles with regard to the magnetic properties.

\begin{figure}
\begin{center}\leavevmode
\centerline{\epsfxsize=9 cm \epsfbox{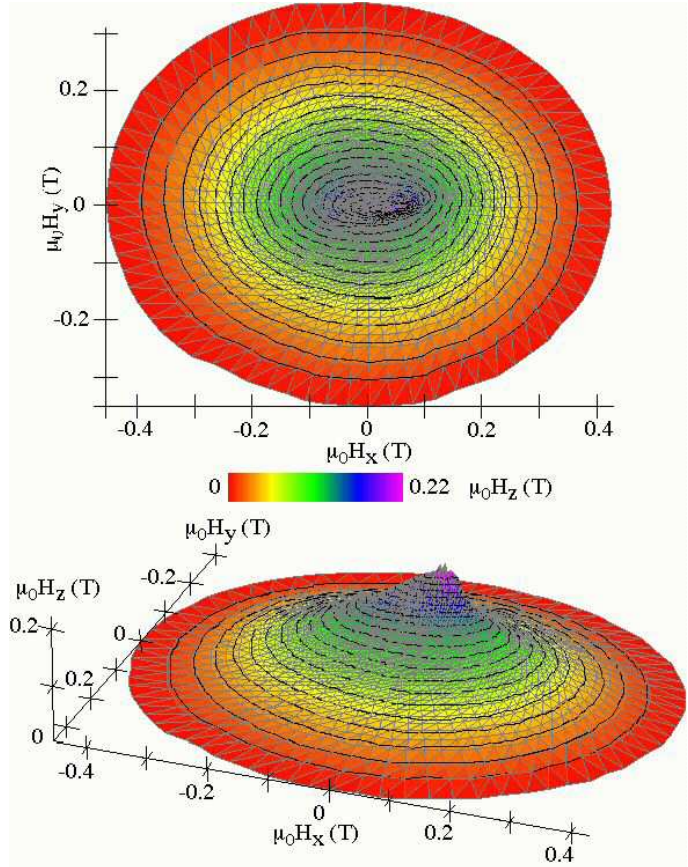}}
\caption{Top view and side view of the experimental 
three-dimensional angular dependence of the switching field 
of a 20 nm Co particle at 35 mK. 
This surface is inversion symmetrical with respect to 
the origin ($H$ = 0).
Continuous lines on the surface are contour lines 
on which $\mu_0 H_z$ is constant.}
\label{part_sciecle}
\end{center}
\end{figure}

\subsection{Measurement technique}
We studied the magnetic properties of individual nanoparticles 
by using planar Nb micro-bridge-DC-SQUIDs (of about 1 
$\mu$m)~\cite{WW_PRL97_Co}. 
In order to place one nanoparticle on the SQUID detector, 
we disperse the particles in ethanol by ultrasonication. 
Then we place a drop of this liquid on a chip of about 
one hundred SQUIDs. When the drop is dry the nanoparticles 
stick on the chip due to Van der Waals forces. 
Only in the case when a nanoparticle falls on a micro-bridge 
of the SQUID loop, the flux coupling between SQUID loop 
and nanoparticle is strong enough for our measurements. 
Finally we determine the exact position and shape of the 
nanoparticles by scanning electron microscopy.

The switching fields of the magnetisation of single Co nanoparticles
were measured by using the {\it cold mode} method  and
the {\it blind mode} method described in 
detail in Ref.~\cite{WW_JAP00_SQUID,Thirion01}.

\begin{figure}
\begin{center}\leavevmode
\centerline{\epsfxsize=9 cm \epsfbox{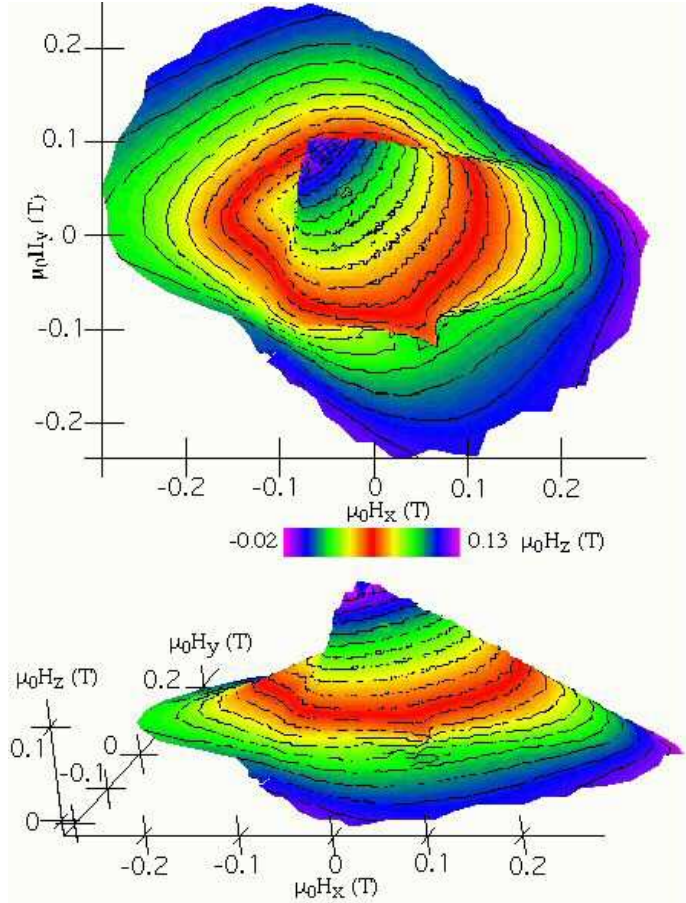}}
\caption{Top view and side view of the experimental 
three-dimensional angular dependence of the switching field 
of a 15 nm Co particle at 35 mK. 
This surface is inversion symmetrical with respect to 
the origin ($H$ = 0).
Continuous lines on the surface are contour lines 
on which $\mu_0 H_z$ is constant.}
\label{Co_fcc_3D}
\end{center}
\end{figure}

\subsection{Results}

The measured angular dependence of switching
fields of nearly all Co nanoparticles revealed a
dominating uniaxial magnetic anisotropy similar to
previous measurements~\cite{WW_PRL97_Co}.
Fig.~\ref{part_sciecle} presents a typical example.
We estimated the shape anisotropy of the typical nearly spherical 
nanoparticles (Fig.~\ref{SEM_Co_1}) and found that
the shape anisotropy constants should be of the order
of magnitude of $10^{4}$ J/m$^{3}$, i.e.
one order of magnitude smaller than the value of the bulk 
f.c.c. cobalt~\cite{Lee90}.
This result suggests that twin boundaries and 
stacking faults (Fig.~\ref{SEM_Co_1}) strongly alter the
cubic crystal symmetry leading to dominating uniaxial anisotropy.

\begin{figure}
\begin{center}\leavevmode
\centerline{\epsfxsize=9 cm \epsfbox{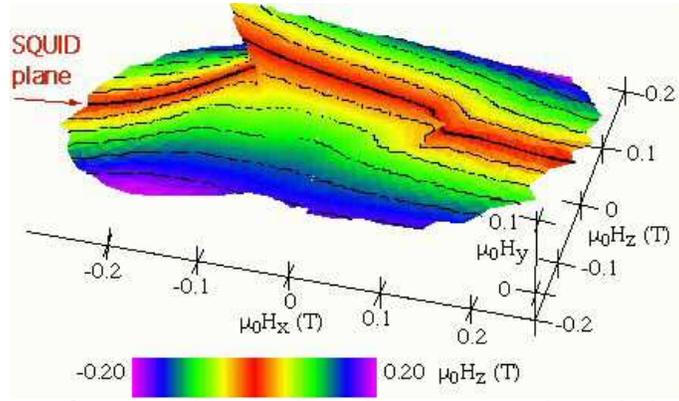}}
\caption{Same data as in Fig. \ref{Co_fcc_3D}.
Continuous lines on the surface are contour lines 
which are parallel to the SQUID plane (defined by $H_z$ = 0).}
\label{plan_SQ}
\end{center}
\end{figure}

\begin{figure}
\begin{center}\leavevmode
\centerline{\epsfxsize=9 cm \epsfbox{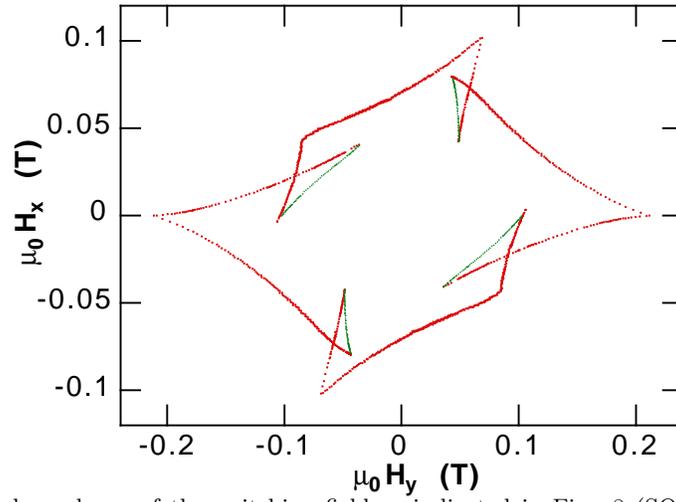}}
\caption{Cut of the angular dependence of the switching field 
as indicated in Fig. \ref{plan_SQ} (SQUID plane). A clear field path dependence 
of the switching field was found (Figs. \ref{ast_fcc} -- \ref{ast_fcc_zoom}) 
which shows a strong influence of 
the cubic crystalline anisotropy.}
\label{Co_fcc_coupe}
\end{center}
\end{figure}

Nevertheless, we could
find few particles which were sufficiently "perfect"
in order to show a more complex switching
field surface and a field path dependence of the switching
field (Fig.~\ref{ast_fcc}) which is the important 
signature of strong higher order terms in the 
potential energy (Eq.~\ref{eq_E}). One example
is presented in Figs.~\ref{Co_fcc_3D}--\ref{Co_fcc_coupe}.
Note that this surface has two easy axes. In addition, 
the hard plan is deformed.
Such a surface can, in principle, be generated by the 
generalised Stoner-Wohlfarth model when taking into
account that the different contributions of 
anisotropies are aligned in an arbitrary way one
with respect to each other.

\begin{figure}
\begin{center}\leavevmode
\centerline{\epsfxsize=9 cm \epsfbox{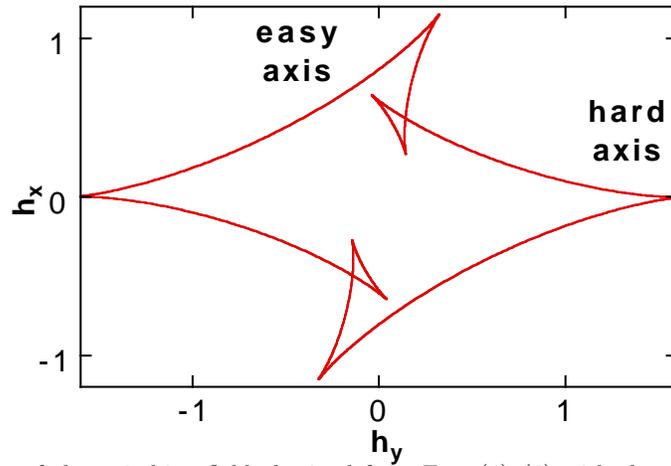}}
\caption{Angular dependence of the 
switching field obtained from 
Eqs.~(\ref{eq_hx})--(\ref{eq_E0_fcc}) with
the same constants as in Fig. \ref{ast_fcc}
but with $\theta_0 = 10^{\circ}$.}
\label{ast_fcc_10}
\end{center}
\end{figure}

In order to understand this point qualitatively, let's
come back to the simple 2D case (Eq. \ref{eq_E0_fcc})
where $\theta_0$ is a constant which allows to turn 
the uniaxial shape anisotropy with respect to the 
cubic anisotropy. Fig. \ref{ast_fcc_10} shows
the angular dependence of the switching field
for a misalignment of $\theta_0 = 10^{\circ}$
which leads to a strong deformation of the curve
in Fig. \ref{ast_fcc}.

\section{Conclusion}

We used the micro-SQUID technique to study 
high quality nanoparticles which were synthesised by arc-discharge.
The angular dependence of switching fields could be
understood in the frame of the simplest classical model 
describing the magnetisation reversal
by uniform rotation. This model were proposed more than 50 years 
ago by N\'eel, Stoner and Wohlfarth. 
The measured critical surface can be considered 
as a geometrical representation of
the magnetic anisotropy of the nanoparticle. 
Nearly all f.c.c. Co nanoparticles revealed a
dominating uniaxial magnetic anisotropy.
This result suggests that twin boundaries and 
stacking faults strongly alter the
cubic magnetocrystalline anisotropy leading to dominating uniaxial anisotropy.
However, few particles were sufficiently "perfect"
in order to show a more complex switching
field surface and a field path dependence of the switching
field. The latter is the important 
signature of the cubic magnetocrystalline anisotropy.

%

\end{document}